\documentclass [12pt]{article}
\usepackage{lscape}                                           %
\usepackage[centertags]{amsmath}
\usepackage{amsfonts} \usepackage{amssymb}\usepackage{eufrak}
\usepackage{graphicx}
\usepackage{a4}
\usepackage{cite}
\usepackage{lscape}

\begin{document}

\begin{titlepage}
\rightline{}
\vskip 3.0cm
\centerline{\LARGE \bf Some Aspects of the T-Duality}
\centerline{\LARGE \bf Symmetric String Sigma Model }

\vskip 1.0cm \centerline{\bf F. Pezzella}
\vskip .6cm
\vskip .4cm 
\centerline{\sl $^c$ Istituto Nazionale di Fisica Nucleare, Sezione di Napoli}
\centerline{ \sl Complesso Universitario di Monte
S. Angelo ed. 6, via Cintia,  80126 Napoli, Italy}
\centerline{ e-mail:  {\sl franco.pezzella@na.infn.it} }

\vskip 0.8cm

\begin{abstract}
A manifestly T-dual invariant formulation of bosonic string theory is discussed here. It can be obtained by making both the usual string compact coordinates and their duals explicitly appear, on the same footing, in the world-sheet action. A peculiarity of such a model is the loss of the local Lorentz invariance which is required to be recovered on-shell. This dictates a constraint on the backgrounds which characterizes the double geometry of the target space. Constant and non-constant backgrounds are considered. In the former case, the local Lorentz constraint implies the geometry of a double torus with an O(D,D) invariance. In the latter, it is shown how and when the O(D,D) invariance still holds and when deformations from it can be implied. Results of the quantization of the free theory are also briefly exhibited.
\end{abstract}

\end{titlepage}

\newpage

\tableofcontents
\vskip 1cm








\section{Introduction and Motivation}

T duality is an old subject in string theory \cite{GP, M}. It implies that in many cases two different geometries for the extra dimensions are physically equivalent. This means that string physics at a very small scale cannot be distinguished from the one at a large scale. 
In the simplest case of compactification of a spatial coordinate $X^{a}$  on a circle of radius $R$, T-duality is encoded by the simultaneous transformations of $R \leftrightarrow \alpha'/R$ and $p_{a} \leftrightarrow w^{a}$, where $p_{a}$ is the quantized momentum of the string and $w^{a}$ is its winding number. Under such transformations, the string coordinate $X^{a}$ along a compact dimension, sum of the left and right movers, 
is transformed into the T-dual coordinate $\tilde{X}_{a}$ 
defined by their difference. 
The winding mode $w^{a}$ plays with respect to $\tilde{X}_{a}$ the same role as $p_{a}$ does with respect to the coordinate $X^{a}$. On a torus $T^{d}$, with strings in the background provided by constant metric $G$ and Kalb-Ramond field $B$, T-duality is described by an $O(d,d; {\mathbb Z})$ 
symmetry. Hence, it would be interesting to extend the standard formulation, based on the
Polyakov action, in a target space with the full duality group $O(D,D)$, where all of the $D$ coordinates are doubled, thus looking for a manifestly T-dual
invariant formulation of string theory \cite{Tsey, Sie, Duff, Hull, 
BT, LP, C, FP2, DGMP, BAND, NIBB}. Introducing both of the coordinates $X^{\mu}$ and $\tilde{X}_{\mu}$ is required. 
If interested in writing down
the complete effective field theory of such generalized sigma-model, one should consider a dependence on both $X^{\mu}$ and $\tilde{X}_{\mu}$ of the fields associated with string states, in particular of the $(G,B)$-background. In this sense,  the gravitational double field theory effective action becomes a double field theory \cite{HZ, AMN, LHZ} of which it is interesting to study symmetries and  properties that could shed light on aspects of string gravity unexplored thus far.

\section{Double String Sigma Model}

The  general string double sigma model we are going to consider is described by the following action \cite{Tsey}:

\begin{eqnarray}
S= - \frac{T}{2} \int d^{2} \sigma \,\, e \left[ C_{ij} \nabla_{0} \chi^{i} \nabla_{1} \chi^{j} + M_{ij} \nabla_{1} \chi^{i} \nabla_{1} \chi^{j} \right] \label{Sdouble}
\end{eqnarray}
where $e^{a}_{\alpha}$ is the zweibein defined on the string world-sheet, $e$ its determinant, the functions $\chi^{i}$ $(i=1, \dots, 2D)$ are the string coordinates in a $2D$-dimensional Riemannian target space, $C_{ij}$ and $M_{ij}$ are symmetric matrices that are, in general, considered to be functions of $\chi^{i}$. Furthermore $\nabla_{a} \chi^{i} = e_{a}^{\,\, \alpha} \partial_{\alpha} \chi^{i}$. This action represents the extension to the case of $2D$ two-dimensional scalar fields of the Floreanini-Jackiw Lagrangians \cite{FJ}, i.e. of the Lagrangians describing the dynamics of the two-dimensional scalar fields $\phi_{\pm}$: 
\begin{eqnarray}
{\cal L}_{\pm} (\phi_{\pm}) = \pm \frac{1}{2} \dot{\phi}_{\pm} \phi^{'}_{\pm} - \frac{1}{2} \phi^{'2}_{\pm} 
\end{eqnarray}
that on shell become, respectively, chiral and antichiral fields, i.e. functions of $\sigma \pm \tau$:
\begin{eqnarray}
\dot{\phi}_{+} = \phi'_{+}  \,\,\,\,\,\,\,\,\  \dot{\phi}_{-} = -\phi'_{-} .
\end{eqnarray}

\subsection{Symmetries, constraints and equations of motion}

The action (\ref{Sdouble}) is symmetric under diffeomorphisms
$\sigma^{\alpha} \rightarrow \sigma'^{\alpha} (\sigma)$ and 
Weyl transformations
$e^{a}_{\,\,\, \alpha} \rightarrow \lambda (\sigma) e^{a}_{\,\,\,\, \alpha}   
$. Instead, it is not manifestly invariant under local Lorentz transformations: $
\delta e^{a}_{\,\,\, \alpha}  = \alpha( \sigma) \epsilon^{a}_{\,\, b} (\sigma) e^{b}_{\, \, \alpha} $
but the theory is required to be locally Lorentz invariant on shell. 
Since the variation of $S$ under such transformations results to be
\begin{eqnarray}
\frac{ \delta S}{\delta e^{a}_{\,\,\, \alpha}} {\delta e^{a}_{\,\,\, \alpha}} = \alpha(\sigma) \frac{e}{2} \epsilon^{a}_{\,\,\, b} t_{a}^{\,\,\, b},  
\end{eqnarray}
this requirement implies $ 
\epsilon^{ab} t_{ab}=0 $ \mbox{with}  $t_{a}^{\,\,\, b} \equiv - \frac{2}{T} \frac{1}{e} \frac{\delta S}{\delta e^{a}_{\,\,\, \alpha}}  e^{b}_{ \,\,\, \alpha}
$
while the Weyl invariance dictates the condition  $\eta^{ab} t_{ab} = 0$ .
 
These local symmetries allow to fix the {\em flat gauge} $
e_{\alpha}^{\,\, a}=\delta_{\alpha}^{\,\, a} .
$ 
The constraint $\epsilon^{ab}t_{ab}=0$ can be rewritten in the following way:
\begin{equation}
 \left[ C_{ij} \partial_{0} \chi^{j} + M_{ij} \partial_{1} \chi^{j} \right] \!\!(C^{-1})^{ik}\!\! \left[ C_{kl} \partial_{0} \chi^{l} + M_{kl} \partial_{1} \chi^{l} ) \right] + \left[ C-MC^{-1}M \right]_{ij}\! \partial_{1} \chi^{i} \,  \partial_{1} \chi^{j} = 0 . \label{constr}
\end{equation} 
 
The equations of motion for $\chi^{i}$ result in:
\begin{eqnarray}
\partial_{1} \left[ C_{ij} \partial_{0} \chi^{j} + M_{ij} \partial_{1} \chi^{j} \right] - \Gamma^{l}_{\,\, ik} C_{lj} \partial_{0} \chi^{j} \partial_{1} \chi^{k}  
- \frac{1}{2} (\partial_{i} M_{jk} ) \partial_{1} \chi^{j} \partial_{1} \chi^{k} = 0 \label{eqmot}
\end{eqnarray}   
where $\Gamma^{l}_{\,\, ik}$ is the Levi-Civita connection constructed out of the matrix $C_{ij}$.
Boundary conditions are given by:
\begin{eqnarray}
\left[ \left( \frac{1}{2} C_{ij} \partial_{0} \chi^{j} + M_{ij} \partial_{1} \chi^{j} \right) \right]^{\sigma=\pi}_{\sigma = 0} = 0.
\end{eqnarray}
\section{Constant background and emerging out of $O(D,D)$}
When $C$ and $M$ are constant \cite{Tsey, MA}, the equations of motion for $\chi^{i}$ drastically simplify into
\begin{equation}
 C_{ij} \partial_{0} \chi^{j} + M_{ij} \partial_{1} \chi^{j}  =0 \,\,\, ,
\end{equation}
after using the further local gauge invariance exhibited by the action under shifts as:
$
\delta \chi^{i} = f^{i}(\tau)$. This causes the constraint (\ref{constr}) to become
$C=MC^{-1}M$. 

After rotating and rescaling $\chi^{i}$, $C$ can always be put in the diagonal form
$
C=  (1, \cdots, 1, -1, \cdots, -1)$
and $\chi^{i}$ becomes $(X^{\mu}_{+},X^{\mu}_{-})$ $(\mu = 1,...,D)$ with $N_{+}=D$ eigenvectors $X^{\mu}_{+}$ associated with the eigenvalue $1$ and  $N_{-}=D$ eigenvectors $X^{\mu}_{-}$ associated with eigenvalues $-1$,  which guarantees the absence of a quantum Lorentz anomaly. 
Hence, $C$ becomes the $O(D,D)$ invariant metric while the condition $C=MCM$ indicates that $M$ has to be an $O(D,D)$ element.
It is possible to make a change of coordinates in the $2D$-dimensional target space: 
$X^{\mu} \equiv \frac{1}{\sqrt{2}}  \left( X^{\mu}_{+} + X^{\mu}_{-} \right) \,\,\, ; \,\,\, \tilde{X}_{\mu} \equiv \delta_{\mu \nu}  \frac{1}{\sqrt{2}} \left( X^{\nu}_{+} - X^{\nu}_{-} \right)$.
The matrix $C$ then becomes off-diagonal:
\begin{center}
$
C_{ij}=-\Omega_{ij}  ~~;~~\Omega_{ij}=\left(\begin{array}{cc}
           0_{\mu \nu} & \mathbb{I}^{ \,\, \nu}_{\mu} \\
           \mathbb{I}^{\mu}_{\,\, \nu}&0^{\mu \nu}\end{array}\right)\label{34}$
\end{center}
while $M$ results in being the {\em generalized metric}:
\begin{eqnarray} 
 M_{ij}= \left( \begin{array}{cc} (G-B\,G^{-1}B)_{\mu \nu} & (B\, G^{-1})_{\mu}^{\,\, \nu}\\
                                       (-G^{-1}\, B)^{\mu}_{\,\,\nu}  & (G^{-1})^{\mu \nu} \end{array}\right)  . \label{generaM}
\end{eqnarray}  

The sigma-model action 
  is invariant under the combined $O(D,D)$ transformations of $\chi^{i}$ {\em and} the matrix of the coupling parameters in $M$:
\begin{eqnarray}
{\cal \chi}'={\cal R} {\cal \chi}~~;~~M'={\cal R}^{-t}M {\cal R}^{-1}~~;~~{\cal R}^{t}\Omega {\cal R} =\Omega~~;~~{\cal R} \in O(D,D).   \nonumber 
\end{eqnarray}
If one considers the duality transformation ${\cal R}=\Omega$ under which $X^{\mu} \rightarrow \tilde{X}_{\mu}$, then the  action, expressed in terms of $X^{\mu}$ and $\tilde{X}_{\mu}$, after such transformation, becomes:
\begin{eqnarray}
S[X, \tilde{X}]= -\frac{T}{2} \int d^{2} \sigma  \left[ \partial_{0} X^{\mu} \partial_{1} \tilde{X}_{\mu} + \partial_{0} \tilde{X}^{\mu} \partial_{1} X_{\mu} 
- (G-B\,G^{-1}B)_{\mu \nu} \partial_{1} X^{\mu} \partial_{1} X^{\nu} \right. \nonumber \\
  -  (B\, G^{-1})_{\mu}^{\,\, \nu} \partial_{1} X^{\mu} \partial_{1} \tilde{X}_{\nu} \nonumber  +  \left. (G^{-1}\, B)^{\mu}_{\,\, \nu} \partial_{1} \tilde{X}_{\mu} \partial_{1} X^{\nu} - (G^{-1})^{\mu \nu} \partial_{1} \tilde{X}_{\mu} \partial_{1} \tilde{X}_{\nu} \right]  \nonumber
  \end{eqnarray}
and exhibits what in string theory is the familiar T-duality invariance, in the presence of backgrounds, i.e. $X \leftrightarrow \tilde{X}$ together with a transformation of the generalized metric given by
 $M'=M^{-1}$, i.e.:
\begin{eqnarray}
G+B \leftrightarrow (G+B)^{-1} \,\,\,\,;\,\,\,\,
G \leftrightarrow (G-BG^{-1}B)^{-1} \,\,\,\,;\,\,\,\,
BG^{-1} \leftrightarrow -G^{-1}B   \,\,\, .
\end{eqnarray}

\subsection{Correspondence to the Standard Formulation}

In order to understand the relationship with the standard formulation, one can integrate over $\tilde{X}_{\mu}$ by eliminating the latter through the use of its equations of motion. In the case of constant $(G,B)$, one gets the standard sigma-model action:
\begin{eqnarray}
S[X] = -\frac{T}{2} \int d^{2} \sigma (\sqrt{G} G^{m m} + \epsilon^{mn}) (G+B)_{\mu \nu} \partial_{m} X^{\mu} \partial_{n} X^{\nu} \nonumber
\end{eqnarray}
which describes the toroidal compactification under proper periodicity conditions on $X$. If, instead, one eliminates $X$ through its equation of motion one obtains the dual model for $\tilde{X}$:
\begin{eqnarray}
S[\tilde{X}] = -\frac{T}{2} \int d^{2} \sigma (\sqrt{G} G^{m n} + \epsilon^{mn}) [(G+B)^{-1}]^{\mu \nu} \partial_{m} \tilde{X}^{\mu} \partial_{n} \tilde{X}^{\nu} \,\,\, . \nonumber
\end{eqnarray}

The generalized model action $S[X, \tilde{X}]$ is therefore a first-order action which interpolates between $S[X]$ and $S[\tilde{X}]$ and is manifestly duality symmetric.

\subsection{Duality Symmetric Free Closed Strings}

From the above formulation it is easy to derive the free dual symmetric action. This corresponds to the case in 	which:

\begin{eqnarray}
 C= -\left( \begin{array}{cc}
           0  & \bf{1} \\
           \bf{1}& 0\end{array}\right) \,\,\, \mbox{and} \,\,\, M = \left(\begin{array}{cc}
           G  & 0 \\
           0& G^{-1}\end{array}\right)  
\end{eqnarray}                   
with $G_{\mu \nu}$ being the {\em flat metric}  in the target space. One gets:
\begin{eqnarray}
 S_{0}=-\frac{1}{4 \pi \alpha'} \int d^{2} \sigma e \left[ \nabla_{0} X^{\mu} \nabla_{1} \tilde{X}_{\mu} +
 \nabla_{0} \tilde{X}^{\mu} \nabla_{1} X_{\mu} 
  - G_{\mu \nu} \nabla_{1} X^{\mu} \nabla_{1} X^{\nu}  - \tilde{G}^{\mu \nu} \nabla_{1} \tilde{X}_{\mu} \nabla_{1} \tilde{X}_{\nu} \right]  \nonumber 
\end{eqnarray}
with  $\tilde{G}^{\mu \nu} = (G^{-1})^{\mu \nu}$, $\nabla_{a} = e_{a}^{\,\,\, \alpha} \partial_{\alpha}$ and $\mu = 1, \cdots, D$.
This is invariant under 
 $X^{\mu} \leftrightarrow \tilde{X}_{\mu}$ together with 
$G_{\mu \nu} \leftrightarrow \tilde{G}^{\mu \nu}$.

The free action $S_{0}$ still describes $D$, not $2D$, scalar degrees of freedom (only the zero mode of $X$ and $\tilde{X}$ are independent on-shell). 
$S_{0}$ can be perturbated by $S_{int} [X, \tilde{X}]$ with the insertion of vertex operators involving both $X$ and $\tilde{X}$. If $S_{int}$ does not depend on $\tilde{X}$ one can integrate $\tilde{X}$ out of the path integral and reproduce the usual results of the standard formulation.

Assuming that strings are compactified on a circle of radius $R$, one should expect that when $R >> \sqrt{\alpha'} $ the relevant interactions are $S_{int}(X)$ ; at intermediate scales  $R \equiv \sqrt{\alpha'}$ the relevant interactions involve both $X$ and $\tilde{X}$ while at $R << \sqrt{\alpha'}$ the relevant interactions are $S_{int}(\tilde{X})$. 

{\em The duality symmetric formulation may be considered as a natural generalization of the standard one at the string scale.}

\section{Non-Constant Backgrounds  }

One can introduce interactions and understand if the local Lorentz constraint  still holds under the form $C=MCM$ in the presence of non-constant backgrounds \cite{Tsey, MA}. 
In the case in which $C=-\Omega$ and $M$ is only $X$-dependent (or only $\bar{X}$-dependent),
then one must take into consideration the contribution coming from the term \cite{FP}
$
 \frac{1}{2} (\partial_{i} M_{jk} ) \partial_{1} \chi^{j} \partial_{1} \chi^{k}
$ in eq. (\ref{eqmot}). The equations of motion for $X^{\mu}$ and $\tilde{X}_{\mu}$ respectively become:

\begin{eqnarray}
\partial_{1} \left[ - \partial_{0} \tilde{X}_{\mu} + (G-BG^{-1}B)_{\mu \nu} \partial_{1} X^{\nu} + (BG^{-1})_{\mu}^{\,\, \nu} \partial_{1} \tilde{X}_{\nu} \right]  = \nonumber \\
 \frac{1}{2} \partial_{1} X^{\nu} \left[ \partial_{\mu} (G-BG^{-1}B)_{\nu \rho} \partial_{1} X^{\rho}  
 + \partial_{\mu} (BG^{-1})^{\nu \rho} \partial_{1} \tilde{X}_{\rho} \right]\nonumber = 0
\end{eqnarray}

and

\begin{eqnarray}
\partial_{1} \left[ - \partial_{0} X^{\mu} + (-G^{-1}B)^{\mu}_{\,\, \nu} \partial_{1} X^{\nu} + (G^{-1})^{\mu \nu} \partial_{1} \tilde{X}_{\nu} \right]  =  \nonumber \\
  \frac{1}{2} \partial_{1} \tilde{X}_{\nu} \left[ \bar{\partial}^{\mu} (-G^{-1} B)^{\nu}_{\,\, \rho} \partial_{1} X^{\rho}  
 + \bar{\partial}^{\mu} (G^{-1})^{\nu \rho} \partial_{1} \tilde{X}_{\rho} \right] = 0 \nonumber
\end{eqnarray}
where $\bar{\partial}^{\mu}$ denotes the derivative with respect to $\tilde{X}_{\mu}$. 

 Also in this case, one can use the further invariance of the theory under   the shifts seen already above and   get for  $X$ the following equation of motion:
\begin{eqnarray}
- \partial_{0} \tilde{X}_{\mu} + (G-BG^{-1}B)_{\mu \nu} \partial_{1} X^{\nu} + (BG^{-1})_{\mu}^{\,\, \nu} \partial_{1} \tilde{X}_{\nu} =0\nonumber \\
- \partial_{0} X^{\mu} + (-G^{-1}B)^{\mu}_{\,\, \nu} \partial_{1} X^{\nu} + (G^{-1})^{\mu \nu} \partial_{1} \tilde{X}_{\nu} = 0 .
\end{eqnarray}  

After substituting the latter equations in the condition $\epsilon_{ab} t^{ab}=0$ in 
eq. (\ref{constr}), which is valid for both constant and non-constant backgrounds,
one can easily see that the off-diagonal structure of $C$ makes the first term vanish in eq. (\ref{constr}) and so one again arrives at the condition $C=MCM$ characterizing the $O(D,D)$ invariance. 
Hence, in this case, the constraint $C=MCM$ is still valid and the expression for $M$ continues to be that in eq. (\ref{generaM})
but now with $X$-dependent $G$ and $B$.  An analogous result is obtained if one considers that $C=-\Omega$ and $M$ is only $\bar{X}$-dependent. 



In the case in which $C$ and $M$ are both non-constant, actually with $C$ and $M$ both depending only on $X$ (or $\bar{X}$), one has to consider, in the equations of motion, also the contribution coming from $-\Gamma^{l}_{\,\, ik} C_{lj} \partial_{0} \chi^{j} \partial_{1} \chi^{k} $.
When rewritten explicitly, this quantity 
vanishes whenever the index $i$ in $ \Gamma^{l}_{\,\, ik} $ runs over the indices of $\tilde{X}_{\mu}$ and therefore it does not give any contribution to the equation of motion of this coordinate.  
One can conclude that the condition $C=MCM$ still holds under the hypothesis that $C$ and/or $M$ are dependent only on $X$ (or only $\tilde{X}$).

When 
both $C$ and $M$ depend on the coordinates $\chi^{i}$, one can think to introduce a  small parameter $\epsilon=\frac{\sqrt{ \alpha'}}{R}$ and to expand $C$ and $M$ up to the second order according to: 
\begin{eqnarray}
C=C_{0} + \epsilon C_{1} + \epsilon^{2} C_{2} \,\,\,\,\,\,;\,\,\,\,\,\,  
M=M_{0} + \epsilon M_{1} + \epsilon^{2} M_{2}  \nonumber 
\end{eqnarray}

By linearizing the condition $\epsilon_{ab}t^{ab}=0$ and the equations of motion for the coordinates, one gets, at the order $\epsilon$, the following expression:
%
\begin{eqnarray}
 (\epsilon_{ab}t^{ab})_{\mbox{on-shell}}  =  -\frac{1}{2} Q_{ij} \partial_{1}\chi^{i} \partial_{1} \chi^{j} + O(\epsilon) \nonumber
\end{eqnarray}
with $Q  =  C_{1} - (C_{0}^{-1} M_{0})^{t}M_{1} - M_{1} (C_{0}^{-1} M_{0}) 
+ (C_{0}^{-1} M_{0})^{t} C_{1} (C_{0}^{-1} M_{0}) = 0. $ 
The above condition can actually be derived by linearizing the condition $C=MCM$. At this order of $\epsilon$, the constraint $C=MCM$ is still valid, being the first term in the expression of the constraint (\ref{constr}) of order $\epsilon^{2}$.
This means that the latter plays a role starting from that order and the contribution coming from it adds to the term proportional to ($C-MCM$) in eq. (\ref{constr}). Starting from the order $\epsilon^{2}$, it seems that the $O(D,D)$ invariance does not hold anymore and one can ask if the deformation is compatible with $O(D,D)$ \cite{FP}.

\section{Quantization of the Double String Model - Non Commutativity}

The Dirac quantization method, necessary because the theory is characterized by primary second-class constraints \cite{DGMP}, implies that $X^{\mu}$ and $\tilde{X}_{\mu}$ behave like non-commuting phase-space type coordinates: 
\begin{eqnarray}  
\left[ X(\tau, \sigma), \tilde{X}(\tau, \sigma') \right] = \frac{i}{T} {\mathbb I} \epsilon (\sigma - \sigma')
\end{eqnarray} \nonumber
with $\epsilon(\sigma) \equiv \frac{1}{2} \left[ \theta(\sigma) - \theta(-\sigma) \right]$.
but it can be shown that their expressions in terms of Fourier modes generate the usual oscillator algebra of the standard formulation \cite{DGMP}.

From this perspective, this non-commutativity may lead to the interpretation of high-energy scattering in the $X$-space as effectively {\em probing} the $\tilde{X}$-space.

\section*{Acknowledgments}

The author would like to deeply thank Olaf Hohm for many and very interesting discussions. Thanks are also due to Hai Lin for providing his feedback on this topic. The author is also grateful to the organizers of the Fourteenth Marcel Grossmann Meeting - MG14 for the stimulating atmosphere they managed to create and for their invitation.

\end{document}